\documentclass[11pt]{article}
\usepackage{moriond,epsfig}

\def\be{\begin{equation}}
\def\ee{\end{equation}}
\def\bea{\begin{eqnarray}}
\def\eea{\end{eqnarray}}

\begin{document}
\vspace*{4cm}
\title{NC COHERENT $\pi^0$ PRODUCTION IN THE MiniBooNE ANTINEUTRINO DATA}

\author{V. T. NGUYEN}

\address{Department of Physics, Columbia University, 538 W 120th ST,\\
New York, NY 10027}

\maketitle\abstracts{
The single largest background to future $\bar{\nu_{\mu}}\rightarrow \bar{\nu_e}$ ($\nu_\mu \rightarrow \nu_e$) oscillation searches is neutral current (NC) $\pi^{0}$ production.  MiniBooNE, which began taking antineutrino data in January 2006, has the world's largest sample of reconstructed $\pi^{0}$'s produced by antineutrinos.  These neutral pions are primarily produced through the $\Delta$ resonance but can also be created through ``coherent production.''  The latter process is the coherent sum of glancing scatters of antineutrinos off a neutron or proton, in which the nucleus is kept intact but a $\pi^{0}$ is created.  A signature of this process is a $\pi^{0}$ which is highly forward-going.  It is advantageous to study coherent production using antineutrinos rather than neutrinos because the ratio of coherent to resonant scattering is enhanced in antineutrino running.  The first measurement of NC coherent $\pi^{0}$ production in the MiniBooNE antineutrino data is discussed here.}

\section{Neutral Current $\pi^0$ Production}

At low energy, neutral current (NC) $\pi^0$'s are produced via two different mechanisms:
\begin{equation}
\bar{\nu}N \rightarrow \bar{\nu} \Delta \rightarrow \bar{\nu}\pi^0 N  \qquad (resonant)
\end{equation}
\begin{equation}
\bar{\nu}A \rightarrow \bar{\nu} A \pi^0  \qquad (coherent)
\end{equation}

\noindent In resonant $\pi^0$ production, a(n) (anti)neutrino interacts with the target, exciting the nucleon into a $\Delta^0$ or $\Delta^+$, which then decays to a nucleon plus $\pi^0$ final state.  In coherent $\pi^0$ production, very little energy is exchanged between the (anti)neutrino and the target.  The nucleus is left intact but a $\pi^0$ is created from the coherent sum of scattering from all the nucleons.  A signature of this process is a $\pi^0$ which is highly forward-going.

\subsection{Why Study NC $\pi^0$ Production?}
NC $\pi^0$ events are the dominant background to $\bar{\nu_{\mu}}\rightarrow \bar{\nu_e}$ ($\nu_\mu \rightarrow \nu_e$) oscillation searches.  A $\pi^0$ decays very promptly into two photons ($\tau_{\pi^0}\sim8$x$10^-17$s), and can mimic a $\bar{\nu_e}$ ($\nu_e$) interaction if only one photon track is resolvable in a detector.

In particular, coherent production is much more challenging to predict theoretically than resonant processes.  Unfortunately, there are currently only two published measurements of the absolute rate of antineutrino NC $\pi^0$ production (Figure 1); the lowest energy measurement reported with 25$\%$ uncertainty at 2 GeV \cite{1}.  There exist no experimental measurements below 2 GeV.  Furthermore, current theoretical models on coherent $\pi^0$ production \cite{2,3,4} can vary by up to an order of magnitude in their predictions at low energy, the region most relevant for (anti)neutrino oscillation experiments.

The analysis presented at the 43rd Rencontres de Moriond represents the first time we are experimentally probing this process in this energy region.  

\begin{figure}[!h]
\begin{center}
  \includegraphics[height=.36\textheight]{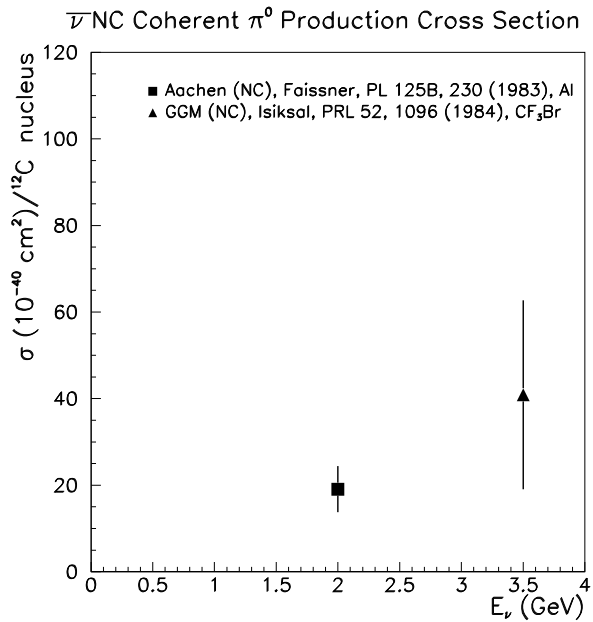}
  \caption{World data on antineutrino NC coherent $\pi^0$ production at low energy.}
\end{center}
\end{figure}

\section{The MiniBooNE Experiment}
The Mini Booster Neutrino Experiment (MiniBooNE) \cite{5}, an experiment at Fermilab designed to measure $\nu_\mu \rightarrow \nu_e$ oscillations, turns out to be very well-suited for $\pi^0$ physics.  Its large, open-volume Cherenkov detector with full angular coverage provides excellent $\pi^0$ identification and containment.  In fact, MiniBooNE has the world's largest samples of NC $\pi^0$ events in interactions with $\sim$1 GeV neutrinos ($\sim$28k) and with $\sim$1 GeV antineutrinos ($\sim$1.7k).  Additional POT has been collected in $\nu$ mode since the MiniBooNE oscillation results \cite{5} with additional POT being collected in $\bar{\nu}$ mode currently. 

\section{Coherent $\pi^0$'s in Terms of $E_\pi(1-\cos{\theta_\pi})$ in $\bar{\nu}$ Mode}
As mentioned before, coherent and resonant $\pi^0$ production are distinguishable by $\cos{\theta_\pi}$, which is the cosine of the lab angle of the outgoing $\pi^0$ with respect to the beam direction.  It turns out that it is even better to study coherent $\pi^0$'s in terms of the pion energy-weighted angular distribution since in coherent events, $E_\pi(1-\cos{\theta_\pi})$ has a more regular shape as a function of momentum, than $\cos{\theta_\pi}$ alone.  Thus, we will fit for the coherent content as a function of the pion energy-weighted angular distribution.  Furthermore, we need to fit this quantity simultaneously with the invariant mass.  This is due to the fact that, in the energy-weighted angular distribution, the resonant and background spectra have similar shapes that are distinct from the forward-peaking coherent spectrum while the resonant and coherent spectra have similar shapes in the invariant mass that are distinct from the background spectrum.

This fit has in fact been done in neutrino mode \cite{6}.  Preliminary fits to the antineutrino data were shown at this conference.  This sample is important because in antineutrino scattering, there is a helicity suppression for most interactions, including resonant production of $\pi^0$'s, but not for coherent production.  Thus, the ratio of coherent to resonant scattering, which is small, is expected to be enhanced in antineutrino running.



\subsection{Preliminary Results}

Preliminary statistics-only fits between MiniBooNE antineutrino data and MC \footnote{RES = NUANCE channels 6,8,13, and 15.  COH = NUANCE channel 96.  BGD = NUANCE channels other than 6,8,13,15, and 96.  MC TOT = RES + COH + BGD.  See Ref. 7 for channel definitions.} are shown (Figures 2 and 3).  The initial MC includes a rescaling of the Rein-Sehgal \cite{8} coherent cross section based on the measurement in Ref. 6.  These studies clearly show evidence for NC coherent $\pi^0$ production, as was the case in neutrino mode.  When coherent $\pi^0$ production is absent from the MC (Figure 4), one can see the poor agreement between data and MC in the lower end of $E_\pi(1-\cos{\theta_\pi})$, where most of the $\pi^0$'s are expected to be produced from coherent scattering.  The agreement improves considerably when coherent $\pi^0$ production is included (Figure 3).  Furthermore, the coherent fraction in both $\nu$ and $\bar{\nu}$ modes is roughly 1.5 times lower than the Rein-Sehgal \cite{8} prediction, which is the most widely used model in (anti)neutrino experiments.  

\begin{figure}[!h]
\begin{center}
  \includegraphics[height=.36\textheight]{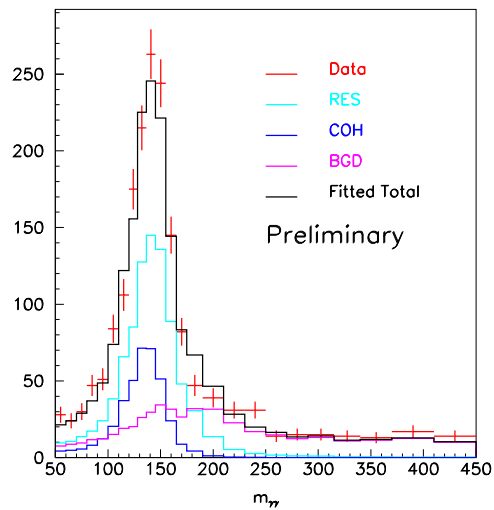}
\vskip -2.0cm
  \caption{Preliminary statistics-only invariant mass fit.}
\end{center}
\end{figure}

\begin{figure}[!h]
\begin{center}
  \includegraphics[height=.36\textheight]{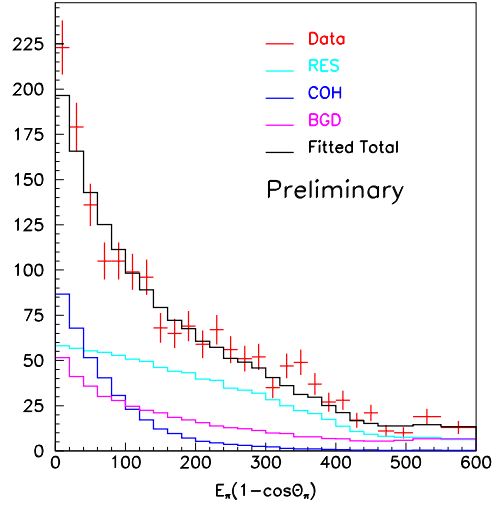}
\vskip -2.0cm
  \caption{Preliminary statistics-only pion energy-weighted angular distribution fit.}
\end{center}
\end{figure}

\begin{figure}[!h]
\begin{center}
  \includegraphics[height=.36\textheight]{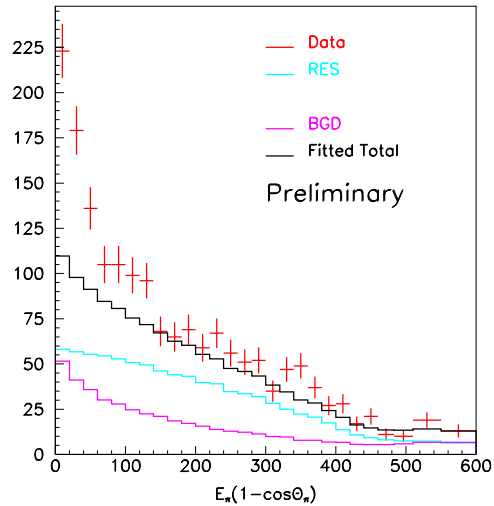}
\vskip -2.0cm
  \caption{Preliminary statistics-only pion energy-weighted angular distribution fit with no coherent contribution in the total MC.}
\end{center}
\end{figure}

\section*{Acknowledgments}
I would like to give many thanks to the MiniBooNE Collaboration and the NSF.

\end{document}